# Exploring Methods for Integrating and Augmenting Multimodal Data to Improve Prognostic Accuracy in Imbalanced Datasets for Intraoperative Aneurysm Occlusion

Parisa Naghdi[1,2], Mohammad Mahdi Shiraz Bhurwani[3], Ahmad Rahmatpour[1,2], Parmita Mondal[1,2], Michael Udin[1,2], Kyle A Williams[1,2], Swetadri Vasan Setlur Nagesh[1,2] Ciprian N Ionita[1,2,3]

[1]Department of Biomedical Engineering, University at Buffalo, Buffalo, NY 14260, [2]Canon Stroke and Vascular Research Center, Buffalo, NY 14203, [3]QAS.AI Inc., Buffalo, NY 14203

## ABSTRACT

**Purpose:** This study investigates the efficacy of a multimodal machine learning framework in predicting treatment outcomes for intracranial aneurysms (IAs), focusing on how hemodynamic alterations captured through angiographic parametric imaging (API) are fused with detailed patient information and disease morphology from digital subtraction angiography (DSA). The research explores various data augmentation strategies to manage dataset imbalances, aiming to enhance the model's accuracy. By integrating deep learning with this comprehensive, fused dataset, the study seeks to significantly improve the precision and predictability of prognostic assessments for IAs.

**Materials and Methods:** Data from 340 patients with treated intracranial aneurysms were analyzed using DSA. We utilized a multimodal approach, integrating quantitative and categorical data—ranging from angiographic parameters at the aneurysm dome to patient demographics like age, gender, and hypertension. Separate deep neural networks were trained for each data type. Pre-decision layers from these networks were concatenated and used as inputs to a final network designed to predict treatment outcomes. Additionally, different deep network architectures were tested for their ability to process and combine these data types effectively. To manage data imbalances, strategies such as random oversampling, undersampling, and Synthetic Minority Over-sampling Technique for Nominal and Continuous (SMOTE-NC) were implemented. The model's performance was evaluated through a 20-split Monte Carlo crossvalidation, focusing on metrics including the area under the receiver operating characteristic (ROC) curve.

**Results:** Our study indicates that augmenting multimodal predictive models for intracranial aneurysm outcomes with techniques like SMOTE/SMOTENC significantly enhances model performance. Intermediate fusion models, in particular, showed notable percentage improvements in predictive accuracy. The effectiveness of other augmentation methods varied, suggesting that the choice of technique is crucial for optimizing results.

**Conclusions:** This study validates the effectiveness of a multimodal machine learning framework in improving the accuracy of predicting treatment outcomes for intracranial aneurysms, with data augmentation techniques substantially enhancing model performance.

**Summary:** This study evaluates a multimodal machine learning framework for predicting treatment outcomes in intracranial aneurysms (IAs). Combining angiographic parametric imaging (API), patient biomarkers, and disease morphology, the framework aims to enhance prognostic accuracy. Data from 340 patients were analyzed, with separate deep neural networks processing quantitative and categorical data. These networks' pre-decision layers were concatenated and inputted into a final predictive network. Various data augmentation strategies, including Synthetic Minority Over-sampling Technique for Nominal and Continuous data (SMOTE-NC), addressed dataset imbalances. Performance metrics, evaluated through Monte Carlo cross-validation, showed significant improvements with augmentation, particularly in intermediate fusion models. This study validates the framework's efficacy in accurately predicting IA treatment outcomes, demonstrating that data augmentation techniques can substantially enhance model performance.

# 1. BACKGROUND

Intracranial aneurysms (IAs) are vascular dilations of the arteries that can rupture and lead to subarachnoid hemorrhage (SAH), one of the most lethal types of strokes. Of those who survive SAH, 66% suffer permanent neurological deficits, underscoring the critical need for accurate diagnosis and management of IAs. [1, 2] Digital subtraction angiography (DSA) remains the standard imaging technique for IA assessment, evaluating both morphology and semiquantitative intra-aneurysmal flow. [3, 4] Recently, enhancements in DSA have enabled angiographic parametric imaging (API), which provides detailed angioarchitecture and flow dynamics estimation by analyzing time-density curves (TDCs) at each pixel to derive parameters such as Mean Transit Time (MTT), Time to Peak (TTP), Peak Height (PH), and Area Under the TDC (AUC). [5, 6] These parameters are important for analyzing flow dynamics in IAs and evaluating changes post-treatment.

API parameters, when analyzed using deep learning techniques, have demonstrated a significant correlation with treatment outcomes of IAs, achieving accuracy levels around 0.8. [7] However, there is potential to further enhance these predictive models. In clinical practice, physicians consider a variety of factors beyond API parameters, such as social factors and aneurysm morphology, which are important in assessing the progression of healing after endovascular surgery. To replicate this multifaceted decision-making process, our study extends the deep learning model to incorporate these additional variables. By employing late and intermediate fusion techniques in multimodal machine learning, we integrate diverse data types, including biomarkers, morphological details, and patient demographic information alongside API parameters. This comprehensive data integration aims to mimic the complex analysis typically performed by clinicians and is expected to significantly improve the precision of the predictive models, providing a more holistic approach to predicting treatment outcomes.

This study employs deep neural networks with multiple hidden layers to explore how different fusion strategies affect the predictive accuracy of IA treatment outcomes, particularly occlusion. Furthermore, considering the imbalanced nature of clinical data, where successfully treated cases far outnumber failures, data augmentation techniques such as oversampling have been implemented. These methods are crucial for training robust models that can generalize well across real-world datasets. By comparing the effectiveness of late and early fusion methods, both with and without oversampling, this research aims to refine predictive algorithms that support better clinical decisions in the treatment of IAs.

# 2. MATERIALS AND METHODS

**2.1 Data Collection**

In this study, patient data collection and subsequent analysis procedures adhered strictly to the protocols approved by the Institutional Review Board at the University at Buffalo, ensuring compliance with ethical standards. We gathered retrospective data from 340 patients who underwent treatment for IAs at the Gates Vascular Institute. These patients were treated using advanced pipeline embolization devices (PED) or coil technologies, with each case having a minimum follow-up period of six months post-treatment to monitor and assess long-term outcomes.

Firstly, we assembled a biomarker dataset incorporating key patient demographics such as age, gender, and race, alongside critical clinical history variables including smoking status, presence of diabetes, hypertension, subarachnoid hemorrhage (SAH) history, body mass index (BMI), and family history of aneurysm. These factors are essential as they can influence both the development and treatment outcomes of intracranial aneurysms. Secondly, our morphology dataset included detailed anatomical data on the aneurysms. This dataset detailed the specific types of aneurysms encountered, such as Cavernous, PCOM (Posterior Communicating Artery), SHA (Superior Hypophyseal Artery), Paraophthalmic, Paraclinoid, Supraclinoid, Petrous, Ophthalmic, PCA (Posterior Cerebral Artery), Dorsal, Cervical, Petrocavernous, Terminus, PICA (Posterior Inferior Cerebellar Artery), and Parasellar. Each type has distinct characteristics that affect treatment strategy and prognosis. Additional data on the aneurysm's physical dimensions, including type, size, the length of the proximal and distal artery segments, and the ratio of the aneurysm dome to neck, were also recorded to better understand their structural implications on blood flow and rupture risk. Lastly, our API dataset, derived from DSA sequences, included parameters such as MTT, TTP, PH, AUC and other relevant API measures. These parameters, obtained from DSAs performed during optimal visualization of the aneurysm dome, provide insights into the blood flow dynamics within the aneurysm, which are critical for predicting rupture risk and planning treatment.

For API data processing, we utilized the QAS.AI Parametric Imaging program (QAS.AI Inc., Buffalo, NY), which facilitated the generation of API maps from DICOM images. We ensured accuracy in our measurements by averaging API feature values within delineated Regions of Interest (ROI) and by manually contouring the aneurysm boundaries. Furthermore, to enhance the reliability of our data against variability in clinical practice, we implemented a two-step normalization method, as proposed by Ionita et al., which was designed to reduce dependency on factors such as injection variability—often influenced by the neurosurgeon's technique—and to adjust for foreshortening effects, where the angiogram's contrast intensity correlates with the length of the X-ray pathway through the anatomical structures.

**2.2 Data fusion**
In this study, predictive models are developed using advanced machine learning techniques to forecast the outcomes of intracranial aneurysm treatments. Data preprocessing involves normalizing both numerical and categorical data to ensure that the inputs for model training are clean and standardized. The first model uses a late fusion approach. The architecture is built using TensorFlow's Keras API. The models consist of multiple deep neural networks that each handle different types of data—such as patient demographics, clinical history, and angiographic parameters—to capture a comprehensive range of features. Each neural network comprises several layers: the input layer is followed by dense layers that use 'LeakyReLU' activation functions for non-linear processing and 'he_normal' as the kernel initializer, which are particularly effective for models with relu and similar activations. Regularization is applied to each layer using L1L2 regularizers to mitigate the risk of overfitting. These networks are then trained to process their respective datasets independently before their outputs are combined. The concatenation of features occurs after the initial individual networks have processed their inputs, allowing the model to integrate diverse data streams effectively. This concatenated output then passes through additional dense layers with 'LeakyReLU' activations to further refine the feature integration, enhancing the model's ability to discern complex patterns across different data types. The final layer of the combined model is a dense layer with a sigmoid activation function, designed to output the probability of treatment success. The model is compiled using the Adam optimizer, with a learning rate carefully set to optimize convergence. The loss function used is binary crossentropy, appropriate for binary classification tasks. Training involves various techniques to optimize model performance, including early stopping to prevent overtraining, model checkpointing to save the best model during training, and a learning rate reduction strategy that adjusts the learning rate when the model's improvement plateaus.

The second model uses an intermediate fusion approach. In this model, preprocessing tasks are streamlined through an advanced pipeline setup, which standardizes numerical features and converts categorical variables using one-hot encoding, ensuring all input data forms are uniformly processed for model training. Unlike the initial model, this refined approach emphasizes more on concatenating features from various data streams at different stages. Here, each feature set derived from individual data sources—such as demographic, clinical, and imaging data—is initially processed through its own dedicated neural network path. These paths include dense layers with configurations tailored to the type of data they process, utilizing activation functions and regularizers designed for optimal performance per data type. After processing, features from these paths are concatenated to form a unified feature set, which then feeds into additional dense layers. This setup allows for a layering strategy where concatenated outputs pass through further neural network layers, enhancing the model's ability to learn complex patterns across combined datasets. Each layer in this architecture is designed to contribute uniquely to handling specific data characteristics before merging, which contrasts with the earlier model where feature integration occurred earlier in the process. This model employs an Adam optimizer like the first, but adjustments in learning rate strategies and the addition of learning rate reduction on plateaus cater to the specifics of the concatenated model, optimizing training efficiency and model accuracy.

For both approaches, we employed early stopping and checkpointing strategies, ensuring that the best model configuration is saved, and overfitting is minimized during training. Overall, both models represent an integrated and data-specific approach to handling diverse datasets, improving upon the unimodal methods, introducing a more complex and finely tuned architecture designed to better mimic clinical decision-making processes.

**2.3 Imbalanced data handling**

In our study, integrating extensive datasets featuring patient characteristics and disease morphology alongside traditional imaging data revealed a significant challenge due to increased problem dimensionality and data complexity. This integration resulted in decreased accuracy of the predictive models due to the exacerbated class imbalances introduced by diverse categorical and binary data types. To address this, we employed several data augmentation strategies to balance our dataset and improve model performance. Specifically, we utilized Synthetic Minority Oversampling Technique (SMOTE) and its extension SMOTE for Nominal Continuous data (SMOTENC). SMOTE helps by synthesizing new examples from the minority class based on existing samples, effectively increasing the representation of under-represented classes in our dataset. SMOTENC adapts this approach to handle mixed data types (both continuous and categorical features) which are prevalent in our datasets due to the inclusion of varied patient demographics and aneurysm characteristics. Additionally, we applied RandomOverSampler to simply duplicate instances of the minority class, aiming for a straightforward increase in their frequency to achieve a more balanced dataset. Furthermore, Adaptive Synthetic Sampling (ADASYN) was used to generate synthetic data focusing more on those regions where the classifier performed poorly. ADASYN adjusts the synthetic sample generation by creating more samples in areas where the model has difficulty learning, thereby improving the overall model training phase.

# 3. RESULTS

In evaluating our predictive models for intracranial aneurysm treatment outcomes, the unimodal approach achieved an AUROC of $0.74\pm0.07$, indicating fair accuracy without augmentation. When employing multimodal fusion techniques, the initial performance was lower (AUROC of $0.68\pm0.1$ for both late and intermediate fusion). However, augmentations such as SMOTE/SMOTENC improved the late fusion model's AUROC to $0.73\pm0.04$ and significantly enhanced the intermediate fusion model to $0.85\pm0.06$, the highest observed performance. The effects of RandomOverSampler and ADASYN varied, with intermediate fusion showing notable declines or moderate improvements (AUROC ranging from $0.52\pm0.07$ to $0.79\pm0.1$), emphasizing that the effectiveness of augmentation strategies can depend heavily on the model structure and data integration approach.

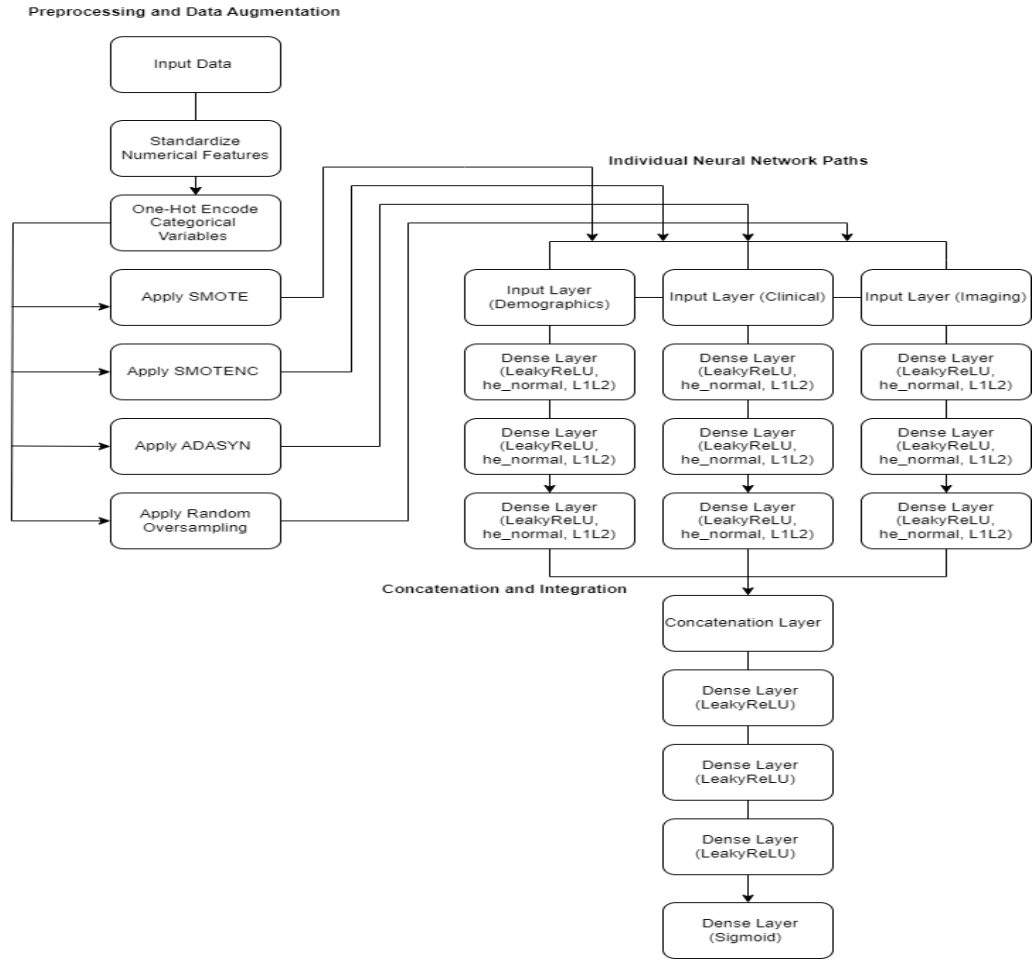

Figure 1. Intermediate fusion model architecture. This architecture is designed to integrate highly-variable datasets, including patient characteristics, disease morphology, and traditional imaging data. This approach aims to enhance predictive model performance despite the increased dimensionality and data complexity. The architecture consists of multiple preprocessing steps, including standardization of numerical features and one-hot encoding of categorical variables, followed by data augmentation techniques such as SMOTE, SMOTENC, random oversampling, and ADASYN to address class imbalance.

## 4. CONCLUSION

This study demonstrated the effectiveness of a multimodal machine learning framework in predicting treatment outcomes for IAs, emphasizing the integration of hemodynamic data with detailed patient information and aneurysm morphology. By applying advanced data augmentation strategies, such as SMOTE/SMOTENC, we significantly improved the precision and predictability of the models. Notably, the use of intermediate fusion techniques coupled with strategic data augmentation resulted in substantial improvements in model accuracy, as evidenced by increases in AUROC values. These findings highlight the potential of combining various data types and sophisticated machine learning techniques to enhance diagnostic tools in a clinical setting. The variability in performance across different augmentation methods also underscores the necessity for careful selection and application of these techniques based on specific model requirements and data characteristics. Moving forward, these insights could guide the development of more refined AI-driven diagnostic and prognostic tools, ultimately contributing to better patient management and treatment outcomes in the field of neurovascular medicine.

## ACKNOWLEDGMENTS

This work is supported by NSF STTR Award # 2111865